\documentclass[aps,prl,reprint,superscriptaddress]{revtex4-1}
\usepackage{hyperref}
\usepackage{graphicx}
\usepackage{enumerate}
\hypersetup{
    pdfnewwindow=true,      
    colorlinks=true,       
    linkcolor=blue,          
    citecolor=blue,        
    filecolor=blue,      
    urlcolor=blue           
}

\usepackage{subfigure}
\usepackage{amsmath}
\usepackage{amssymb}
\usepackage{amsfonts}

\usepackage{color}

\def\be{\begin{equation}}
\def\ee{\end{equation}}

\begin{document}

\title{Observational Constraints on Multi-messenger Sources of Gravitational Waves and High-energy Neutrinos}

\author{Imre Bartos}
\email[]{ibartos@phys.columbia.edu}
\affiliation{Department of Physics, Columbia University, New York, NY 10027, USA}
\author{Chad Finley}
\affiliation{Oskar Klein Centre \& Dept. of Physics, Stockholm University, SE-10691 Stockholm, Sweden}
\author{Alessandra Corsi}
\affiliation{LIGO Laboratory, California Institute of Technology, Pasadena, CA 91125, USA}
\author{Szabolcs M\'arka}
\affiliation{Department of Physics, Columbia University, New York, NY 10027, USA}

\begin{abstract}

It remains an open question to what extent many of the astronomical sources of intense bursts of electromagnetic radiation are also strong emitters of non-photon messengers, in particular gravitational waves (GWs) and high-energy neutrinos (HENs). Such emission would provide unique insights into the physics of the bursts; moreover some suspected classes, e.g. choked gamma-ray bursts, may in fact only be identifiable via these alternative channels.  Here we explore the reach of current and planned experiments to address this question.  We derive constraints on the rate of GW and HEN bursts per Milky Way equivalent (MWE) galaxy based on independent observations by the initial LIGO and Virgo GW detectors and the partially completed IceCube (40-string) HEN detector. We take into account the blue-luminosity-weighted distribution of nearby galaxies, assuming that source distribution follows the blue-luminosity distribution. We then estimate the reach of joint GW+HEN searches using advanced GW detectors and the completed km$^{3}$ IceCube detector to probe the joint parameter space. We show that searches undertaken by advanced detectors will be capable of detecting, constraining or excluding, several existing models with one year of observation.

\end{abstract}

\maketitle

Gravitational-wave (GW) astronomy, as well as high-energy neutrino (HEN) observations are entering a new and promising era with newly constructed detectors, providing unprecedented opportunities to observe these astrophysical messengers, opening new windows onto the universe. GW observatories \cite{LIGO0034-4885-72-7-076901,VIRGO0264-9381-23-19-S01,GEO0264-9381-19-7-321,LCGT0264-9381-27-8-084004} are being built and upgraded to second generation detectors. Several HEN detectors \cite{IceCubeAhrens2004507,Antares} have reached their design sensitivities and will be further upgraded in the near future \cite{Avrorin2011S13}.

GWs and HENs can originate from a number of common sources. Plausible sources include gamma-ray bursts (GRBs), core-collapse supernovae (CCSNe), soft gamma repeaters as well as microquasars \cite{ET}. For a joint GW+HEN analysis, the most interesting sources are those which have little or no detected electromagnetic (EM) emission.
With the near-completion of the first searches for multi-messenger GW+HEN sources \cite{1742-6596-243-1-012002}, it is important to examine the projected science reach of such searches, as well as how
it relates` to independent GW and HEN measurements. This can support and guide the theoretical work necessary to gain a better understanding of future observational results.

In this paper we interpret and combine previously published and independent GW and HEN observational results, to derive the first joint constraints on the rates of GW+HEN sources.
We first discuss constraints from individual HEN and GW searches, and then combine these to derive upper limits on GW+HEN sources. We finally estimate projected constraints on GW+HEN sources with future detectors and joint GW+HEN searches.

\begin{figure}
\begin{center}
\resizebox{0.47\textwidth}{!}{\includegraphics{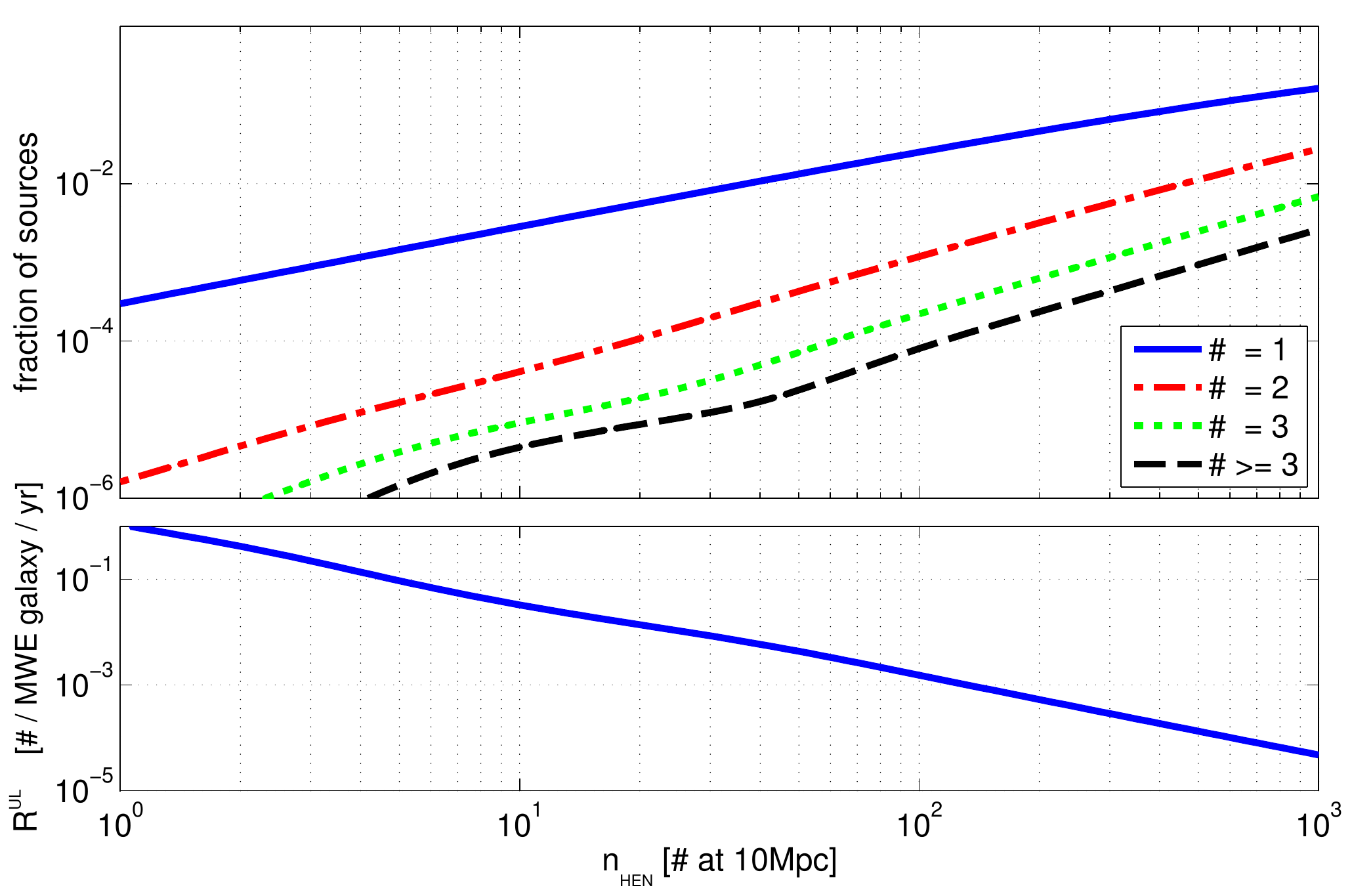}}
\end{center}
\caption{(Color online) \textbf{Top:} fraction of neutrino-emitting sources within 1 Gpc which would be detected with 1, 2, 3, or $\geq3$ neutrinos, as a function of $n_{\textsc{hen}}$ (the mean number of detected neutrinos from a source at 10 Mpc) for a detector with northern sky coverage (e.g. IceCube). Only sources are considered that emit neutrinos towards the Earth. \textbf{Bottom:} source population upper limit $R^{\textsc{ul}}$ as a function of $n_{\textsc{hen}}$, assuming a beaming factor of $f_b=14$, and considering only the northern sky. }
\label{figure:HEN}
\end{figure}

\emph{Upper limits from neutrino observations}--- 
Abbasi \emph{et al.} \cite{2011arXiv1104.0075T} searched for transient point sources with the partially constructed IceCube detector in its 40-string configuration (hereafter IceCube-40) for over 1 year.  The search covered the northern sky and emission time-scales ranging from seconds to months; no evidence for transient sources was found.  The number of neutrinos needed for detection on each time-scale was determined.  Using here the conservatively large time window of 500~s for HEN emission from GRBs by Baret \emph{et al.} \cite{Baret20111}, three spatially coincident neutrinos within this time window would have been sufficient for a ($5\sigma$) discovery.
(We note that even with the higher event rate of IceCube-86, three coincident neutrinos will remain a highly unlikely outcome from the background).
We therefore estimate the source population upper limit as the maximum source rate that has $\lesssim90\%$ probability to result in at least one occurrence of $\geq3$ coincident neutrinos in a time window of 500~s during a 1-year measurement.

We model the source population as following the blue-luminosity distribution of galaxies \cite{1993A&A...273..383C}: (i) for up to 40 Mpc, we take the blue-luminosity distribution given in the Gravitational Wave Galaxy Catalog \cite{0264-9381-28-8-085016} (we note that any incompleteness in the galaxy catalog makes
our upper-limits conservative.); (ii) for larger distances (up to 1~Gpc) we adopt the homogenous blue-luminosity density determined by Blanton \emph{et al.} \cite{0004-637X-592-2-819}; (iii) we assume that IceCube is uniformly sensitive to sources in the northern sky only, which is a reasonable approximation of the detector's directional sensitivity \cite{2011arXiv1104.0075T}. Our upper-limits are calculated as a function of $n_{\textsc{hen}}$, defined as the average number of detected HENs from a source at 10~Mpc (e.g. \cite{chokedfromreverseshockPhysRevD.77.063007}).

The results provided here assume that each source has the same intrinsic neutrino brightness (limits based on a fixed average brightness are conservative compared to those using any other brightness distribution), and account for beaming of the HEN emission.
For a source with intrinsic brightness $n_{\textsc{hen}}$ at distance $r$, the probability that $\geq3$ neutrinos will be detected from it is
\begin{equation}
p(n\geq3|r,n_{\textsc{hen}})=1-F\left(2|(10\mbox{ Mpc}/r)^2 n_{\textsc{hen}}\right),
\end{equation}
where $r$ is the source distance, $F$ is the Poisson cumulative distribution function, and $n$ is the number of detected neutrinos from the source. Therefore for galaxy $i$ with blue luminosity L$_{\textsc{b}}^{(i)}$ at a distance $r_i$, the average number $\widehat{N}_i$ of sources which are discovered (i.e. have $\geq3$ detected neutrinos) will be
\begin{equation}
\widehat{N}_i(R,T) = p(n\geq3|r_i,n_{\textsc{hen}})\cdot R / f_b \cdot T\cdot L_{\textsc{b}}^{(i)}/L_{\textsc{b}}^{\mbox{\textsc{mw}}},
\end{equation}
where $R$ is the source rate [number of sources per year per Milky Way equivalent (MWE) galaxy (w.r.t. blue luminosity)], $f_b$ is the HEN beaming factor of the source, $T$ is the duration of the measurement ($\approx1$ year \cite{2011arXiv1104.0075T}), and $L_{\textsc{b}}^{\mbox{\textsc{mw}}}$ is the blue luminosity of the Milky Way. The $90\%$ confidence source population upper limit $R^{\textsc{ul}}$ will be the upper limit that satisfies $2.3 \geq \sum_i \widehat{N}_i(R^{\textsc{ul}},T)$, i.e.
\begin{equation}
R^{\textsc{ul}}(n_{\textsc{hen}}) = \frac{2.3f_bL_{\textsc{b}}^{\mbox{\textsc{mw}}}}{T\sum_{\delta_i \geq0}p(n\geq3|r_i,n_{\textsc{hen}})L_{\textsc{b}}^{(i)}},
\end{equation}
where the sum is over all galaxies with declination $\delta_i \geq 0$. For $r>40$ Mpc where we consider a homogeneous matter distribution, the summation is substituted with an integral. Figure~\ref{figure:HEN} (top) shows the fraction of HEN sources as a function of $n_{\textsc{hen}}$. In the lower plot, population upper limits for HEN sources are shown, taking into account the sources' HEN beaming factor $f_b$. As mildly relativistic jets from CCSNe and low-luminosity (LL) GRBs are expected to make up a significant portion of HEN sources of interest \cite{HENAndoPhysRevLett.95.061103,chokedfromreverseshockPhysRevD.77.063007,ET}, we adopt $f_b=14$ corresponding to LL-GRB beaming factor obtained in \cite{Liang2007}.

\begin{figure}
\begin{center}
\resizebox{0.47\textwidth}{!}{\includegraphics{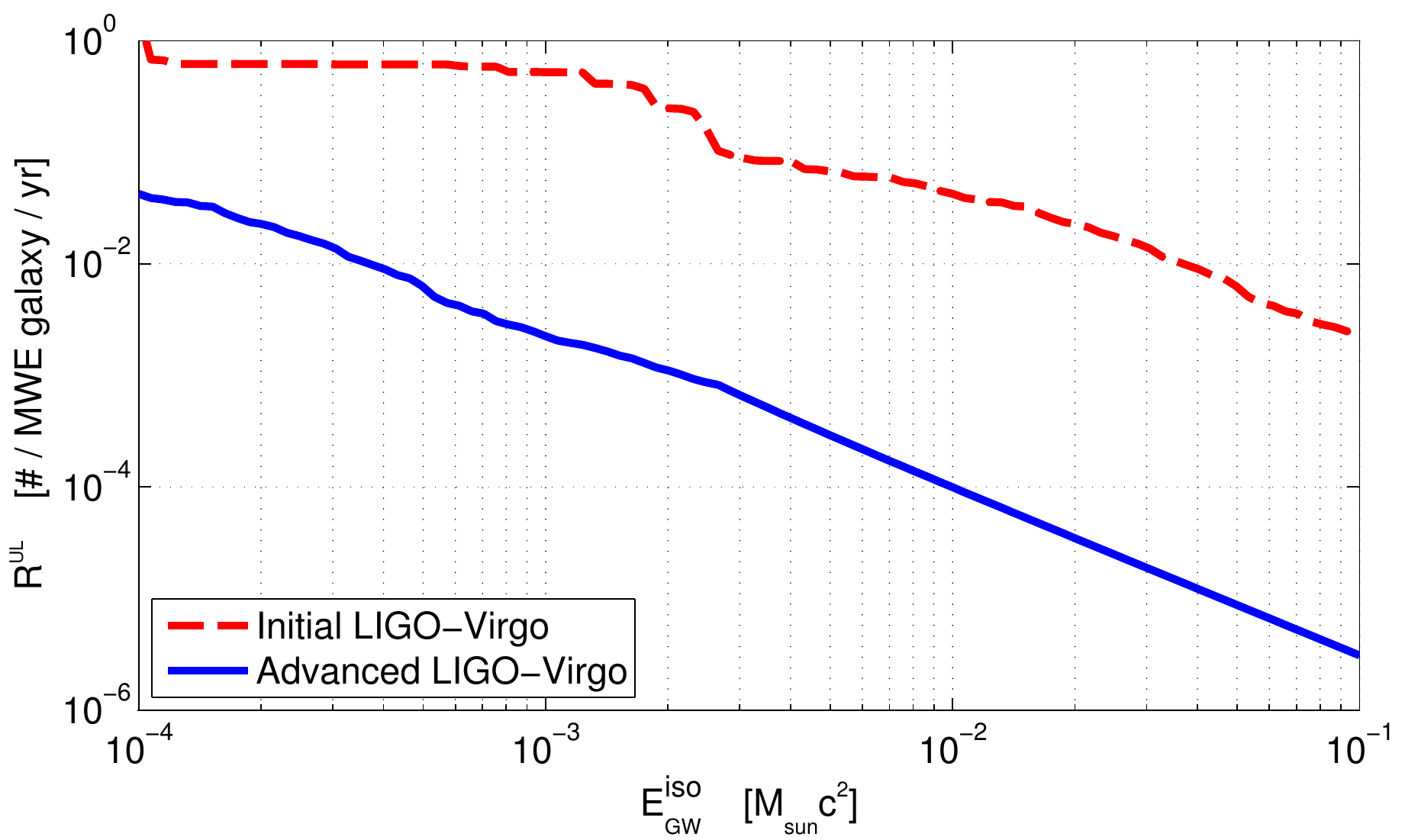}}
\end{center}
\caption{(Color online) Source population upper limits as functions of the sources' GW emission in isotropic-equivalent energy $E_{\textsc{gw}}^{iso}$.  \textbf{Dashed red:} observational limits with initial LIGO-GEO-Virgo \cite{PhysRevD.81.102001}. \textbf{Solid blue}: projected limits for the advanced LIGO-GEO-Virgo GW detectors in the event of non-detection.} \label{figure:GW}
\end{figure}

\emph{Upper limits from gravitational-wave observations}--- 
We use the limits obtained by the latest GW all-sky burst search by Abadie \emph{et al.} \cite{PhysRevD.81.102001}. We consider their result for sine-Gaussian GW waveform in the sensitive band of the GW detectors ($\sim150$~Hz).

Abadie \emph{et al.} report no detection using the initial LIGO-GEO-Virgo detectors \cite{LIGO0034-4885-72-7-076901,GEO0264-9381-19-7-321,VIRGO0264-9381-23-19-S01}, and set a frequentist 90$\%$ confidence upper limit of $\approx0.5\times(10^{-2}M_\odot c^2/E_{\textsc{gw}}^{iso})^{3/2}$/year/Mpc$^3$, or 2.0 detectable events per year, on the population of the considered GW bursts. Here we interpret this result through introducing a GW horizon distance $D^{\textsc{gw}}(E_{\textsc{gw}}^{iso})$, within which any GW bursts with $E_{\textsc{gw}}^{iso}$ energy would have been greater than the loudest background event of the measurement. This is a reasonable approximation of the detection efficiency for the obtained efficiencies in \cite{PhysRevD.81.102001}. The above result yields $D^{\textsc{gw}}(E_{\textsc{gw}}^{iso})~=~7.8\cdot~(E_{\textsc{gw}}^{iso}/10^{-2}M_\odot~c^2)^{1/2}$~Mpc. We thus derive a galaxy-based GW source population upper limit as a function of $E_{\textsc{gw}}^{iso}$, using the blue-luminosity-weighted distribution of galaxies as described in (i)-(ii) above:
\begin{equation}
R^{\textsc{ul}}(E_{\textsc{gw}}^{iso}) = \frac{2.0L_{\textsc{b}}^{\mbox{\textsc{mw}}}}{\sum_{r_i\leq D^{\textsc{gw}}}L_{\textsc{b}}^{(i)}}\cdot\frac{1}{\mbox{year}}.
\end{equation}
Here, we assumed that each GW source emits the same amount of GW energy (limits based on uniform GW emission are conservative compared to those using any other emission distribution). We estimate the achievable population upper limit for the advanced LIGO-Virgo GW detector network by assuming a $\sim10\times$ increase in sensitivity compared to the network of initial detectors, with similar measurement duration. Results are shown in Figure \ref{figure:GW}.

\begin{figure}
\begin{center}
\resizebox{0.47\textwidth}{!}{\includegraphics{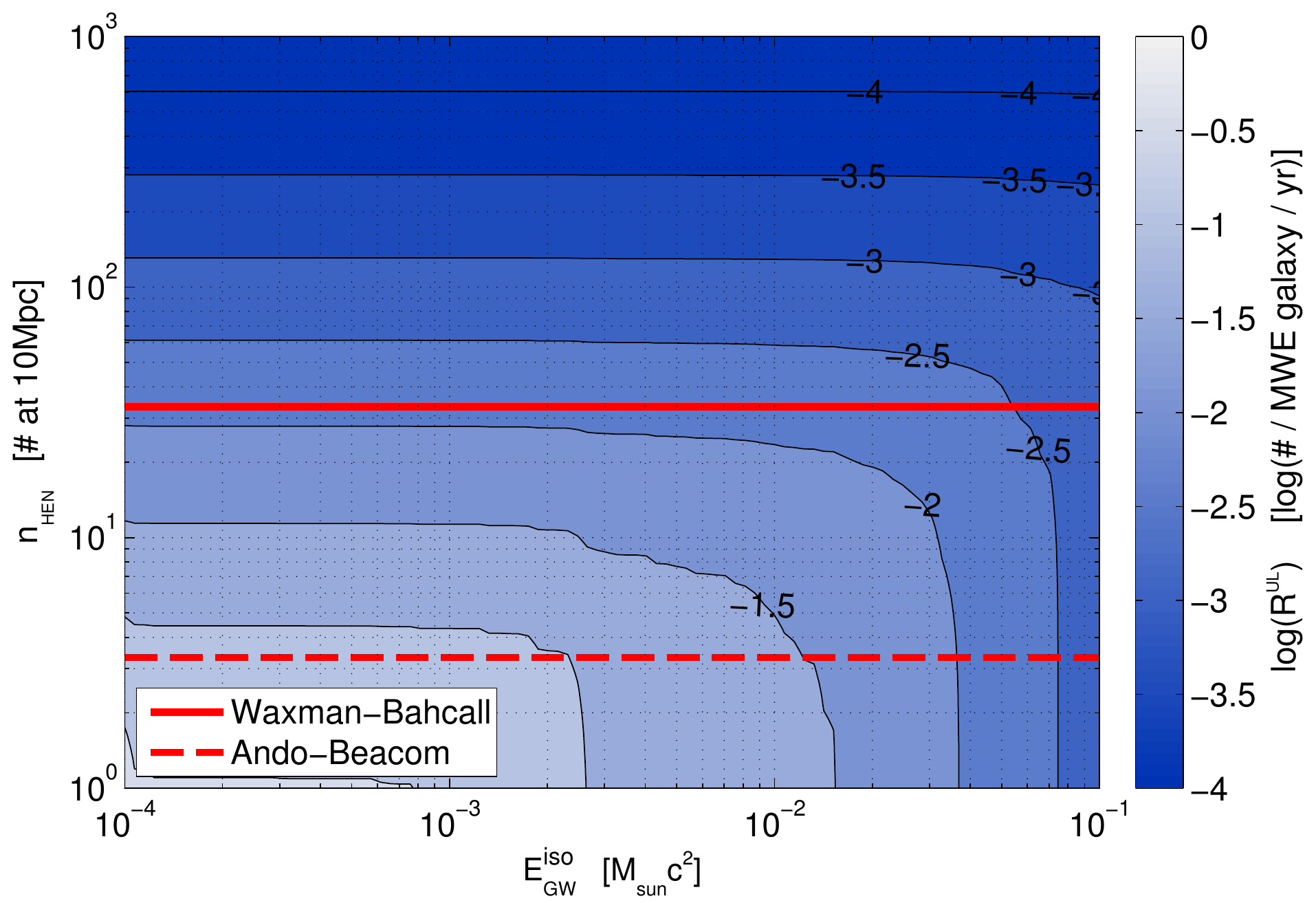}}
\resizebox{0.47\textwidth}{!}{\includegraphics{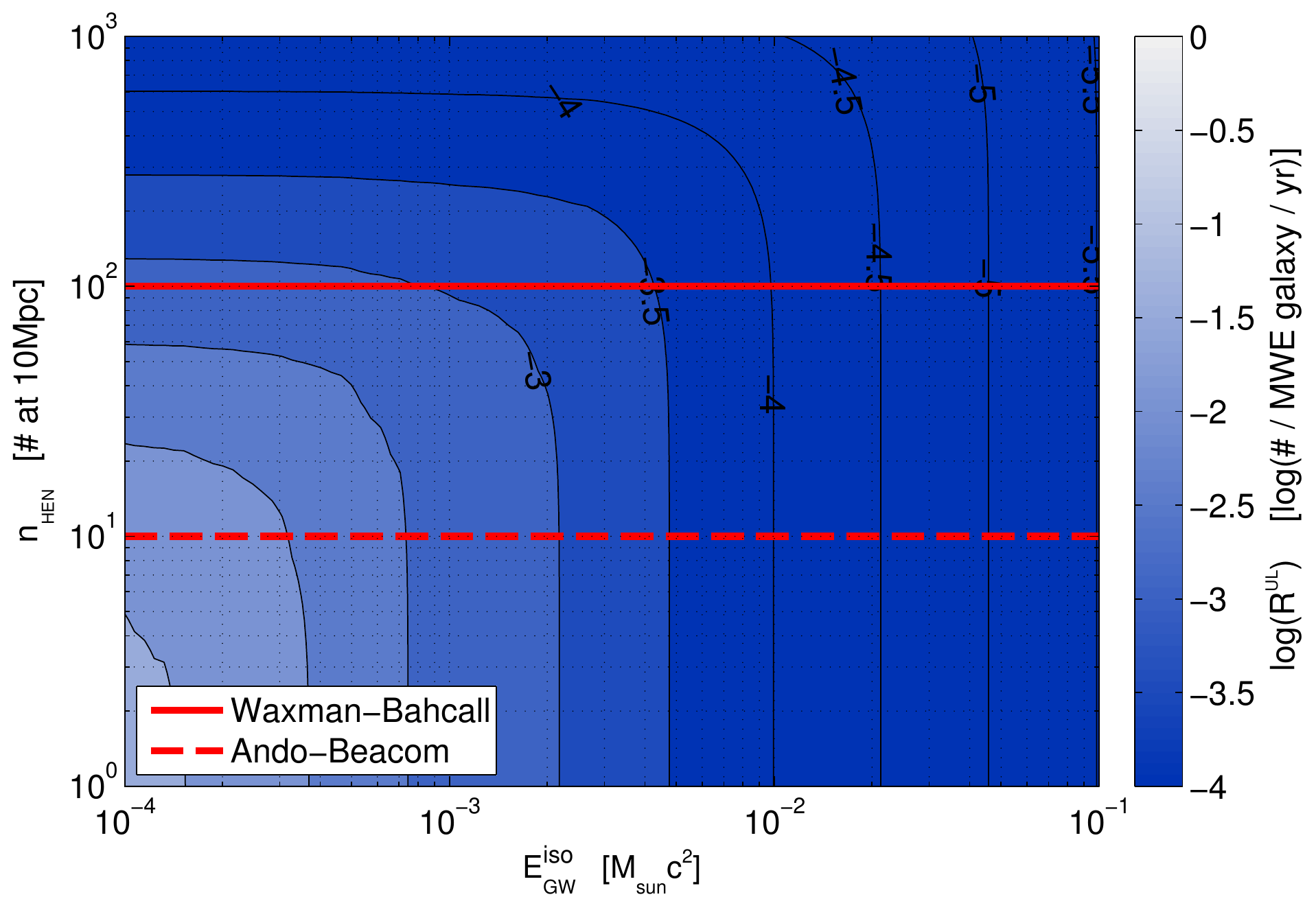}}
\end{center}
\caption{(Color online) GW+HEN source population upper limits based on the statistical combination of independent GW and HEN measurements. \textbf{Top:} observational results for measurements with the initial LIGO-GEO-Virgo GW detectors \cite{PhysRevD.81.102001} and the IceCube-40 HEN detector \cite{2011arXiv1104.0075T}. \textbf{Bottom:} projected results for 1-year observations with advanced LIGO-Virgo and IceCube-86. The limits shown assume a HEN beaming factor of 14. \textbf{Horizontal lines:} expected HEN rate from the Waxman-Bahcall \cite{waxmanbachall} (solid) and Ando-Beacom \cite{HENAndoPhysRevLett.95.061103} (dashed line) models, scaled to the IceCube-40 (top) and IceCube-86 (bottom) detector configurations.} \label{figure:GWHEN}
\end{figure}

\emph{Joint GW+HEN population upper limits}--- 
Individual GW and HEN observations can be combined to determine a GW+HEN source population upper limit in the $E_{GW}^{iso}$--$n_{\textsc{hen}}$ parameter space. In Figure~\ref{figure:GWHEN} (top) we provide GW+HEN population upper limits based on the statistical combination of current observational results from \emph{independent} GW and HEN measurements. We obtain a joint observational upper limit by considering that, on average, less than 2.3 GW+HEN bursts occur within $D^{\textsc{gw}}$ or have $\geq3$ detected HENs per year (this is a $>90\%$ confidence upper limit, since the GW and HEN measurements were longer than one year). The observational GW+HEN upper limit for a source population proportional to the blue-luminosity-weighted galaxy distribution will therefore be
\begin{equation}\begin{split}
&R^{\textsc{ul}}(E_{\textsc{gw}}^{iso},n_{\textsc{hen}}) = \\
&\frac{2.3L_{\textsc{b}}^{\mbox{\textsc{mw}}}\cdot\mbox{year}^{-1}}{\frac{1}{f_b}\sum_{\{r_i> D^{\textsc{gw}}, \delta_i \geq0\}}p(n\geq3|r_i,n_{\textsc{hen}})L_{\textsc{b}}^{(i)} + \sum_{\{r_i\leq D^{\textsc{gw}}\}}L_{\textsc{b}}^{(i)}}.
\label{equation:GWHEN}
\end{split}
\end{equation}
Figure~\ref{figure:GWHEN} (top) also compares the obtained upper limits to two HEN emission models (\cite{waxmanbachall,HENAndoPhysRevLett.95.061103}). As the theoretical estimates in \cite{waxmanbachall,HENAndoPhysRevLett.95.061103} are provided for km$^3$ scale detectors, we convert them to estimates for IceCube-40 by conservatively assuming that the sensitivity of IceCube-40 is one-third that of IceCube-86.

Similarly to the above observational results, we also calculate the projected GW+HEN population upper limits based on the statistical combination of projected results from independent, 1 year long measurements with advanced LIGO-GEO-Virgo and IceCube-86. Results are shown in Figure \ref{figure:GWHEN} (bottom).

\begin{figure}
\begin{center}
\resizebox{0.47\textwidth}{!}{\includegraphics{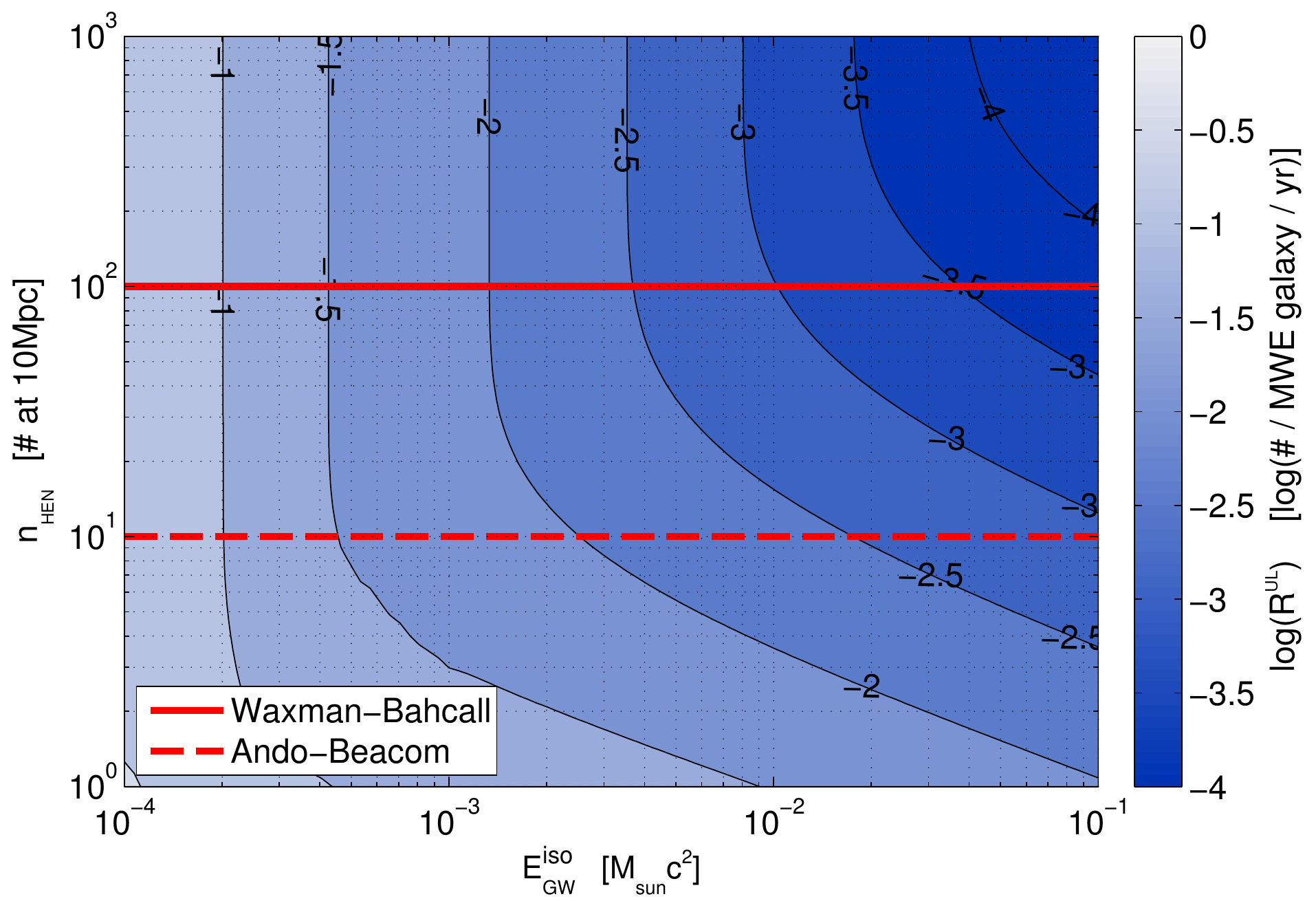}}
\end{center}
\caption{(Color online) Projected GW+HEN source population upper limits for a joint analysis of one-year of observations with advanced LIGO-Virgo and IceCube-86. Results are given as functions of source emission parameters $E_{GW}^{iso}$ (GW emission in isotropic-equivalent energy) and $n_{\textsc{hen}}$ (average number of detected neutrinos from a source at 10 Mpc). \textbf{Horizontal lines:} expected HEN rate from the Waxman-Bahcall \cite{waxmanbachall} (solid) and Ando-Beacom \cite{HENAndoPhysRevLett.95.061103} (dashed line) models. The limits shown assume a HEN beaming factor of 14.} \label{figure:ALL}
\end{figure}

We now estimate the projected population upper limits for GW+HEN sources obtainable with a \emph{joint} GW+HEN search, considering a 1-year measurement with the advanced LIGO-Virgo and IceCube-86 detectors. We consider an event candidate to be the coincidence of 1 GW trigger and 1 HEN. While we might detect more than 1 HEN from some sources, the fraction of such sources is small (see Figure \ref{figure:HEN}), therefore we conservatively omit multi-HEN sources. For the joint search we define a horizon distance $D^{\textsc{gwhen}}(E_{\textsc{gw}}^{iso})$, such that a joint GW+HEN event with 1 detected HEN and GW energy $E_{\textsc{gw}}^{iso}$, within $D^{\textsc{gwhen}}$ would be more significant than the (anticipated) loudest background event.
We estimate $D^{\textsc{gwhen}}$ to be the same as the exclusion distance of the externally triggered search for GW bursts by Abbott \emph{et al.} \cite{0004-637X-715-2-1438}, who obtained a median exclusion distance of $D\sim12\mbox{ Mpc}\cdot (E_{\textsc{gw}}^{iso}/10^{-2}M_{\odot}c^2)^{1/2}$ with GW emission in the most sensitive band of LIGO-Virgo. Such comparison to externally triggered GW searches is a reasonable approximation if the joint search has $O(1)$ chance overlaps of background GW and HEN events (which can be controlled by adjusting the event selection threshold). For the joint GW+HEN search the estimated source population upper limit $R^{\textsc{ul}}$ will be
\begin{equation}\begin{split}
R^{\textsc{ul}}(E_{\textsc{gw}}^{iso},n_{\textsc{hen}}) = \frac{2.3f_bL_{\textsc{b}}^{\mbox{\textsc{mw}}}}{T\sum_{\{r_i \leq D^{\textsc{gwhen}}, \delta_i \geq0\}}p(n\geq1|r_i,n_{\textsc{hen}})L_{\textsc{b}}^{(i)}}.
\end{split}
\end{equation}
The estimated population upper limits for a GW+HEN search are shown in Figure \ref{figure:ALL} for advanced LIGO-Virgo and IceCube-86.

\emph{Discussion.}--- 
To compare our results to emission models, we consider SNe with mildly relativistic jets (SNe with jets) as promising GW+HEN emitters, whose observational rate upper limit has been about $\lesssim1\%$ of the total SN Ib/c rate \cite{2010Natur.463..513S} ($\sim3\times10^{-4}$/year/MWE galaxy).

All-sky population upper limits with IceCube-86 are projected to exclude sources at the rates of SNe with jets, for $n_{\textsc{hen}}\gtrsim 300$ (Fig. \ref{figure:HEN}). This is comparable to the Waxman-Bahcall flux \cite{waxmanbachall}, which is, in the specific case of GRBs, already constrained by HEN searches \cite{PhysRevLett.106.141101}. Upon non-detection, advanced GW detectors are projected to exclude sources at current limits on rates of SN with jets, for $E_{\textsc{gw}}^{iso} \gtrsim 5\times10^{-3} M_\odot c^2$ (Fig. \ref{figure:GWHEN}, bottom panel), excluding the suspended accretion model \cite{PhysRevD.69.044007}, and constraining the accretion disk fragmentation model \cite{1538-4357-579-2-L63}.

The above estimates assume that the fraction of SNe with jets is probed
by radio observations. It has been proposed, however, that mildly
relativistic jets may be much more common, but completely choked (bright
in neutrinos, dark in gamma-rays and radio) \cite{chokedfromreverseshockPhysRevD.77.063007,HENAndoPhysRevLett.95.061103}. The nearby
core-collapse supernova rate is high enough to allow testing these models
soon.
Upon non-detection, IceCube-86 is projected (Fig. \ref{figure:HEN}) to exclude sources at SN rates with $n_{\textsc{hen}}\gtrsim 12$, a level comparable to the emission expected from SN jets by Ando and Beacom ($n_{\textsc{hen}} \approx 10$; \cite{HENAndoPhysRevLett.95.061103}),
or emission through reverse shocks in mildly relativistic jets ($n_{\textsc{hen}} \lesssim7$; \cite{chokedfromreverseshockPhysRevD.77.063007}). Moreover, as evident from Fig. \ref{figure:GW}, advanced GW detectors are projected to exclude sources at SN rates with average
GW emission of $E_{\textsc{gw}}^{iso} \gtrsim 2\times10^{-4} M_\odot c^2$, excluding suspended accretion as well as accretion disk fragmentation.

We obtain projected population constraints with a joint GW+HEN search (Fig. \ref{figure:ALL}) that can be more restrictive in some regions of the parameter space than individual searches if the GW horizon distance $D^{\textsc{gw}}$ of the joint search is $\gtrsim2.4$ times greater ($\sim f_b^{1/3}$) than $D^{\textsc{gw}}$ of individual GW searches.

The authors thank Zsuzsa M\'arka and Eric Thrane for their valuable comments, and Peter M\'esz\'aros for his encouragement. The work of CF was generously supported by the Swedish Research Council (VR), under contract no. 622-2010-383 and through the Oskar Klein Centre. The Columbia Experimental Gravity group is grateful for the generous support from Columbia University in the City of New York and from the National Science Foundation under cooperative agreement PHY-0847182.
LIGO was constructed by the California Institute of Technology and Massachusetts Institute of Technology with funding from the National Science Foundation under cooperative agreement PHY-0757058.
This paper has document number LIGO-P1100098. 

\bibliographystyle{h-physrev}

\end{document}